\documentclass{nature}
\usepackage{graphicx}
\usepackage{bbold}
\usepackage{bm}
\usepackage{amsmath}
\usepackage{amsfonts}
\usepackage{amssymb}
\usepackage{braket}
\usepackage{units}
\usepackage{mathrsfs}
\usepackage{svg}

\begin{document}

\title{Higgs mode in a strongly interacting fermionic superfluid}

\author{A. Behrle$^{1,3}$, T. Harrison$^{1,3}$, J. Kombe$^2$, K. Gao$^1$, M. Link$^1$, J.-S. Bernier$^2$, C. Kollath$^2$, and M. K{\"o}hl$^1$}
\maketitle

\begin{affiliations}
\item Physikalisches Institut, University of Bonn, Wegelerstrasse 8, 53115 Bonn, Germany
\item HISKP, University of Bonn, Nussallee 14-16, 53115 Bonn, Germany
\item Contributed equally.
\end{affiliations}

\textbf{Higgs and  Goldstone modes are  possible collective modes of an order parameter upon spontaneously breaking  a continuous symmetry. Whereas the low-energy Goldstone (phase) mode is always stable, additional symmetries are required to prevent the Higgs (amplitude) mode from rapidly decaying into low-energy excitations. In high-energy physics, where the Higgs boson\cite{Higgs1964} has been found after a decades-long search, the stability is ensured by Lorentz invariance. In the realm of condensed--matter physics, particle--hole symmetry can play this role\cite{Littlewood1981} and a Higgs mode has been observed in weakly-interacting superconductors\cite{Sooryakumar1980,Matsunaga2013,Sherman2015}. However, whether  the Higgs mode is also stable  for strongly-correlated superconductors  in which particle--hole symmetry is not precisely fulfilled or whether this mode becomes overdamped has been subject of numerous discussions\cite{Pekker2015,Podolsky2011,Scott2012,Barlas2013,Liu2016,Han2016}. Experimental evidence is still lacking, in particular owing to the difficulty to excite the Higgs mode directly. Here, we observe the Higgs mode in a strongly-interacting superfluid Fermi gas. By inducing a periodic modulation of the amplitude of the superconducting order parameter $\Delta$,  we observe an excitation resonance at frequency $2\Delta/h$. For strong coupling, the peak width broadens and eventually the mode disappears when the Cooper pairs turn into tightly bound dimers signalling the eventual instability of the Higgs mode.}

Spontaneous symmetry breaking occurs when an equilibrium state exhibits a lower symmetry than the corresponding Hamiltonian describing the system. The system then spontaneously picks one of the energetically degenerate choices of the order parameter and due to the specific energy landscape this process is accompanied by new collective modes. The typical picture, which exemplifies spontaneous symmetry breaking, uses a Mexican-hat shaped energy potential (see Figure 1a) that suggests the emergence of two distinct collective modes: the gapless ``Goldstone mode'', which is associated with long-wavelength phase fluctuations of the order parameter, and an orthogonal gapped mode, the ``Higgs mode'', which describes amplitude modulations of the order parameter. While Goldstone modes, such as phonons, appear necessarily when continuous symmetries are broken, stable Higgs modes are scarce, since decay channels might be present. The best-known example of a Higgs mode appears in the Standard Model of particle physics where this mode gives elementary particles their mass \cite{Higgs1964}.

In the non-relativistic low-energy regime usually encountered in condensed-matter physics, the existence of a stable Higgs mode cannot be taken for granted\cite{Pekker2015}. However, under certain conditions, other symmetries, such as particle-hole symmetry, can play the role of Lorentz invariance and induce a stable Higgs mode. A notable example of a low-energy particle-hole symmetric theory hosting a stable Higgs mode is the famous Bardeen-Cooper-Schrieffer (BCS) Hamiltonian describing weakly interacting superconductors \cite{Littlewood1981,Littlewood1982}. Evidence for the Higgs mode has been found in conventional BCS superconductors \cite{Sooryakumar1980,Matsunaga2013,Sherman2015}. However, experimental detections have been solely indirect as the Higgs mode does not couple directly to electromagnetic fields owing to the gauge invariance required for its existence.  The far-reaching importance of the Higgs mode  is further illustrated by its observation  in a variety of specially tuned systems such as antiferromagnets \cite{Ruegg2008}, liquid $^3$He \cite{Halperin1990}, ultracold bosonic atoms near the superfluid/Mott-insulator transition \cite{Bissbort2011,Endres2012}, spinor Bose gases \cite{Hoang2016}, and Bose gases strongly coupled to optical fields \cite{Leonard2017}. In contrast, weakly-interacting Bose-Einstein condensates do not exhibit a stable Higgs mode \cite{Pekker2015,Liu2016,Han2016}.

In recent years, research has focused on advanced materials exhibiting superconductivity far beyond the conventional BCS description, such as cuprates, pnictides, and the unitary Fermi gas. Many of these materials are characterized by strong fermionic correlations.  Even though in this context the existence of a Higgs mode has been the topic of theoretical debates \cite{Podolsky2011,Scott2012,Barlas2013,Liu2016,Han2016}, experimental evidence for the Higgs mode in systems exhibiting strong correlations between fermions is still absent. 

Here, we spectroscopically excite  the Higgs mode in a superfluid  Fermi gas in the crossover between a weakly-interacting BCS superfluid and a Bose-Einstein condensate (BEC) of strongly-coupled dimers (Fig.~1). We induce a periodic modulation of the amplitude of the superconducting order parameter $\Delta$ and find an excitation resonance near twice the superconducting gap value. On the BCS side, the spectroscopic feature agrees with the theoretical expectation of the Higgs mode. On the BEC side of the crossover, we find strong broadening beyond the predictions of BCS theory and, eventually, the disappearance of the mode as predicted for a weakly-interacting BEC \cite{Pekker2015,Liu2016,Han2016}. 

\begin{figure}
 \includegraphics[width=\columnwidth,clip=true]{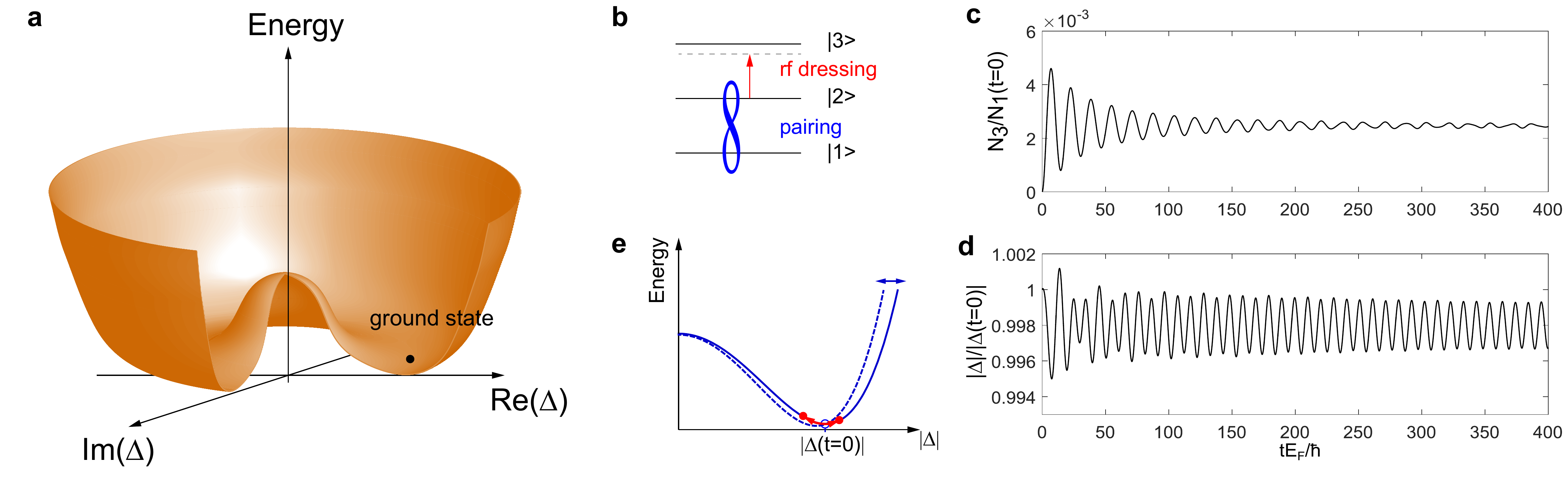}
 \caption{Principle of the Higgs mode excitation. (a) Mexican hat potential of the free energy as a function of real and imaginary part of the complex order parameter $\Delta$. The equilibrium state order parameter takes spontaneously one of the values at the energy minima. (b) We employ radiofrequency dressing of the paired superfluid by off-resonant coupling to an unoccupied state $\ket{3}$. This results in a periodic modulation of both the occupation of the state $\ket{3}$ (c) and the superconducting gap (d). Shown are numerical simulations for a coupling constant $1/(k_Fa)=-0.6704$, $\hbar \Omega_{R} = 0.0353E_{F}$ and $\hbar \delta = -0.3247E_{F}$. (e) By adjusting the modulation  frequency, we achieve an excitation of the Higgs  mode in the Mexican hat. }
\end{figure}

Our measurements are conducted in an ultracold quantum gas of $\sim 4\times 10^6$ $^6$Li atoms prepared in a balanced mixture of the two lowest hyperfine states $\ket{1}$ and $\ket{2}$ of the electronic ground state. The gas is trapped in a harmonic potential with frequencies of $(\omega_x,\omega_y,\omega_z)=2\pi \times (91,151,235) \,$Hz and is subjected to a homogeneous magnetic field, which is varied in the range of $740-1000$\,G in order to tune the s-wave scattering length $a$ near the Feshbach resonance located at 834\,G. This results in an adjustment of the interaction parameter of the gas in the range of $-0.8\lesssim 1/(k_Fa)\lesssim 1$, i.e. across the whole BCS/BEC crossover. The Fermi energy in the center of the gas is $E_F\simeq h\times (34\pm 3)$\,kHz at each of the considered interaction strengths and sets the Fermi wave vector $k_F=\sqrt{8 \pi^2 m E_F/h^2}$, where $m$ denotes the mass of the atom and $h$ is Planck's constant.

Excitation of the Higgs mode requires a scheme which couples to the amplitude of the order parameter rather than creating phase fluctuations or strong single-particle excitations. Previous theoretical proposals\cite{Yuzbashyan2006,Scott2012,Hannibal2015} for exciting the Higgs mode in ultracold Fermi gases have focused on a modulation of the interaction parameter $1/(k_Fa)$, however, experimentally only single-particle excitations have been observed from such a modulation\cite{Greiner2005}. We have developed a novel excitation scheme employing a radiofrequency (rf) field dressing the state $\ket{2}$ with the initially unoccupied hyperfine state $\ket{3}$ thereby modulating the pairing between the $\ket{1}$ and $\ket{2}$ states, see Figure 1b and c. Previous experiments investigating ultracold gases with rf spectroscopy\cite{Chin2004,Ketterle2007,Stewart2008,Feld2011} have focused on studying  single-particle excitations. To this end, there, the duration of the rf pulse $\tau$ had been chosen shorter than the inverse of the Rabi frequency $\Omega_R$, such that the spectra could be interpreted in the weak-excitation limit using Fermi's golden rule. In contrast, here, we employ an rf drive  far red-detuned from single-particle resonances in the interacting many-body system and in the long-pulse limit $\Omega_R\tau \gg 1$, in order to couple to the amplitude of the order parameter. To illustrate this, consider first an isolated two-level system of the $\ket{2}$ and the $\ket{3}$ state coupled by a Rabi frequency $\Omega_R$ with detuning $\delta$ from the resonance. The occupation probability of the atoms in the $\ket{2}$ state is $p_{\ket{2}}=1-\Omega_R^2/{\Omega_R^\prime}^2 \sin^2(\Omega_R^\prime t/2)$, i.e. the continuous rf dressing leads to a time-periodic modulation of the occupation of the $\ket{2}$-state with the effective Rabi frequency $\Omega_R^\prime=\sqrt{\Omega_R^2+\delta^2}$. 

\begin{figure}
 \includegraphics[width=.5\columnwidth,clip=true]{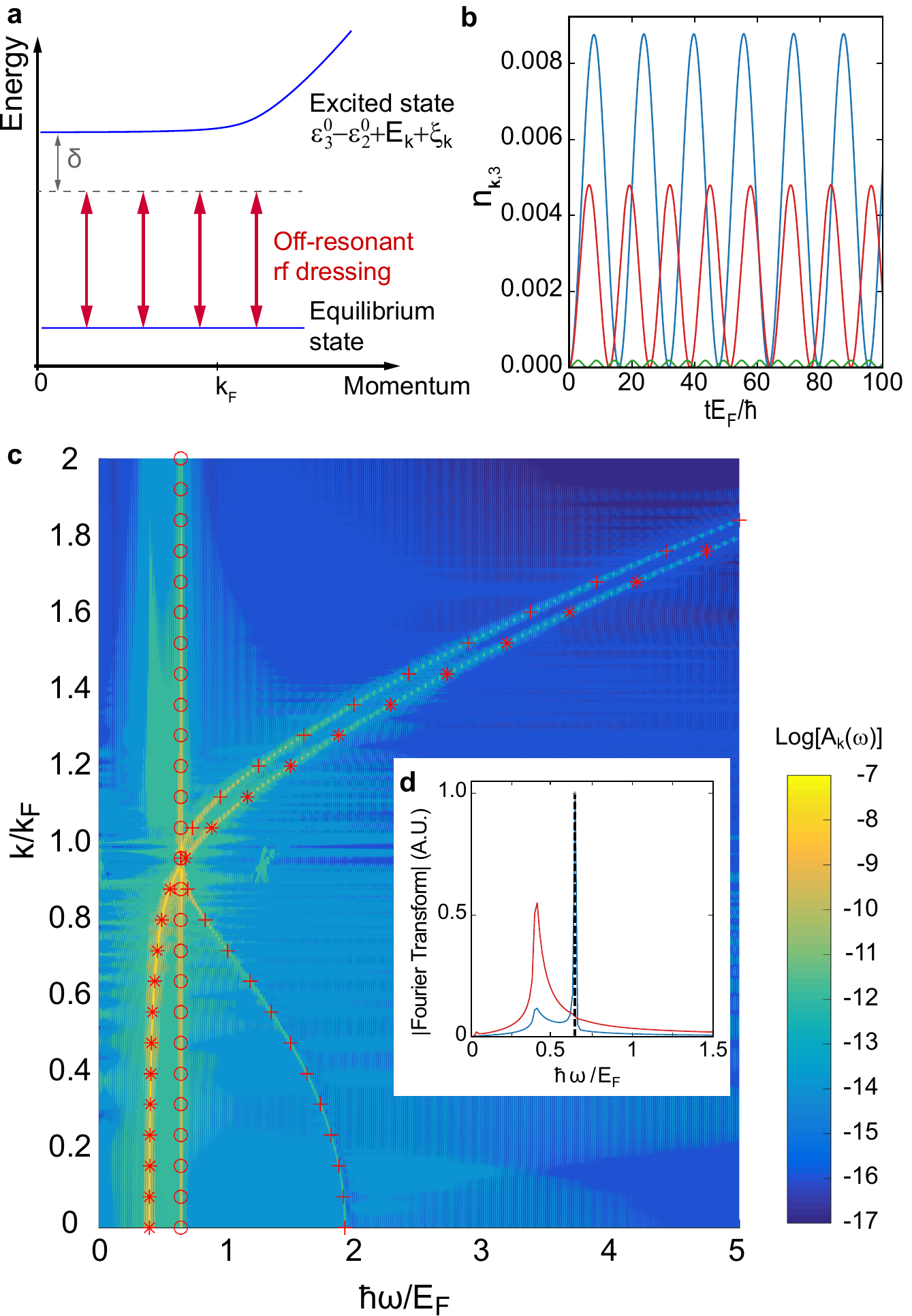}
 \caption{Illustration of the excitation scheme for one modulation frequency. (a) The radiofrequency field is red-detuned from the single-particle excitation of the interacting system. It creates an off-resonant excitation to the state $\ket{3}$  with a varying detuning for different momenta. (b) Time-evolution of the momentum-resolved occupation of the $\ket{3}$ state with momentum $\bm{k}$ for a fixed value of $\frac{1}{k_{F}a} = -0.63$, a Rabi frequency $\hbar \Omega_{R} = 0.038 E_{F}$, and a detuning $\hbar \delta = -0.34 E_{F}$. Blue: $|\bm{k}|/k_F=0$, red: $|\bm{k}|/k_F=0.8$, green: $|\bm{k}|/k_F=1.1$. (c) Spectral weight of the momentum-resolved gap $A_{\textbf{k}}(\omega)$ (see Methods). The circles indicate the Higgs mode, the stars mark the response to the modulation frequency and the crosses indicate the quasiparticle excitations at $2E_k$. The position of the star at $k=0$ approximately represents the effective modulation frequency for the chosen parameters. (d) Fourier spectra (momentum integrated) of the occupation of the $\ket{3}$ state (red) and $|\Delta|$ (blue). The dashed line is the expected location of the Higgs mode at $2|\Delta|$. Subplots b, c, and d are for the same driving and detuning parameters.}
\end{figure}

In the many-body problem of the BCS/BEC crossover, the situation is complicated by the dispersion of the (quasi-) particles and the presence of interactions. In particular, a continuum of excitations typically occurs above the energy of the lowest single-particle excitation to state $\ket{3}$ (see Figure 2a). Deep in the BCS regime, the continuum of excitations is related to the different momentum states  and the excitation scheme can be approximated by coupling  each occupied momentum state  of the BCS quasi-particles in level $\ket{2}$ to the corresponding momentum state in state $\ket{3}$ since the rf dressing transfers negligible momentum. The effective Rabi frequency $\Omega_{R,k}^\prime=\sqrt{\Omega_R^2+\delta_k^2}$ and therefore the excitation probability becomes momentum dependent by the many-body detuning $\hbar \delta_k= \hbar \delta-E_{k}-\xi_k$, where  $\xi_k$ is the single-particle dispersion, and $E_k=\sqrt{\xi_k^2+|\Delta|^2}$ is the quasi-particle dispersion and $\Delta$ is the s-wave superconducting order parameter. A red-detuned rf drive, as employed here, avoids resonant coupling to the single-particle excitations, however, still modulates off-resonantly the occupation of the excited states as shown in Figure 2b. 

The  off-resonant periodic modulation of the occupation of the state $\ket{2}$ with controllable  frequency  $\Omega_{R,k}^\prime$ induces a modulation of the amplitude of the order parameter $|\Delta|$ (Fig. 1e, for details see Methods) and hence couples directly to the Higgs mode.  To illustrate this mechanism, we  numerically solve the minimal set of coupled equations of motion describing the evolution of the order parameter in the presence of an rf coupling to state $\ket{3}$ (see Methods and Supplementary Figure). We see that the Fourier spectrum of $|\Delta|$ for one modulation frequency displays -- aside from a response corresponding to the modulation frequencies $\Omega_{R,k}^\prime$ -- a sharp peak at the gap value $2|\Delta|$, see Figure 2d. By the momentum-resolved representation (Figure 2c) we identify this peak to be dominated by  the Higgs excitation with a momentum-independent dispersion. The amplitude of the latter is maximum  when the averaged effective drive frequency $\hbar \bar{\Omega}_{mod} \approx 2|\Delta|$ in accordance with its expected frequency.  The Higgs mode is a collective mode of the system and even for the harmonically trapped gas exhibits a unique frequency. Numerical studies in the BCS limit have shown that in harmonically trapped systems, the Higgs mode should occur at the frequency of twice the superconducting gap evaluated at the maximum density of the gas \cite{Scott2012,Bruun2002,Korolyuk2011, Korolyuk2014,Tokimoto2017} and hence we use this as our reference for the value of the gap in order to compare with theory and other experiments.

\begin{figure}
 \includegraphics[width=.9\columnwidth]{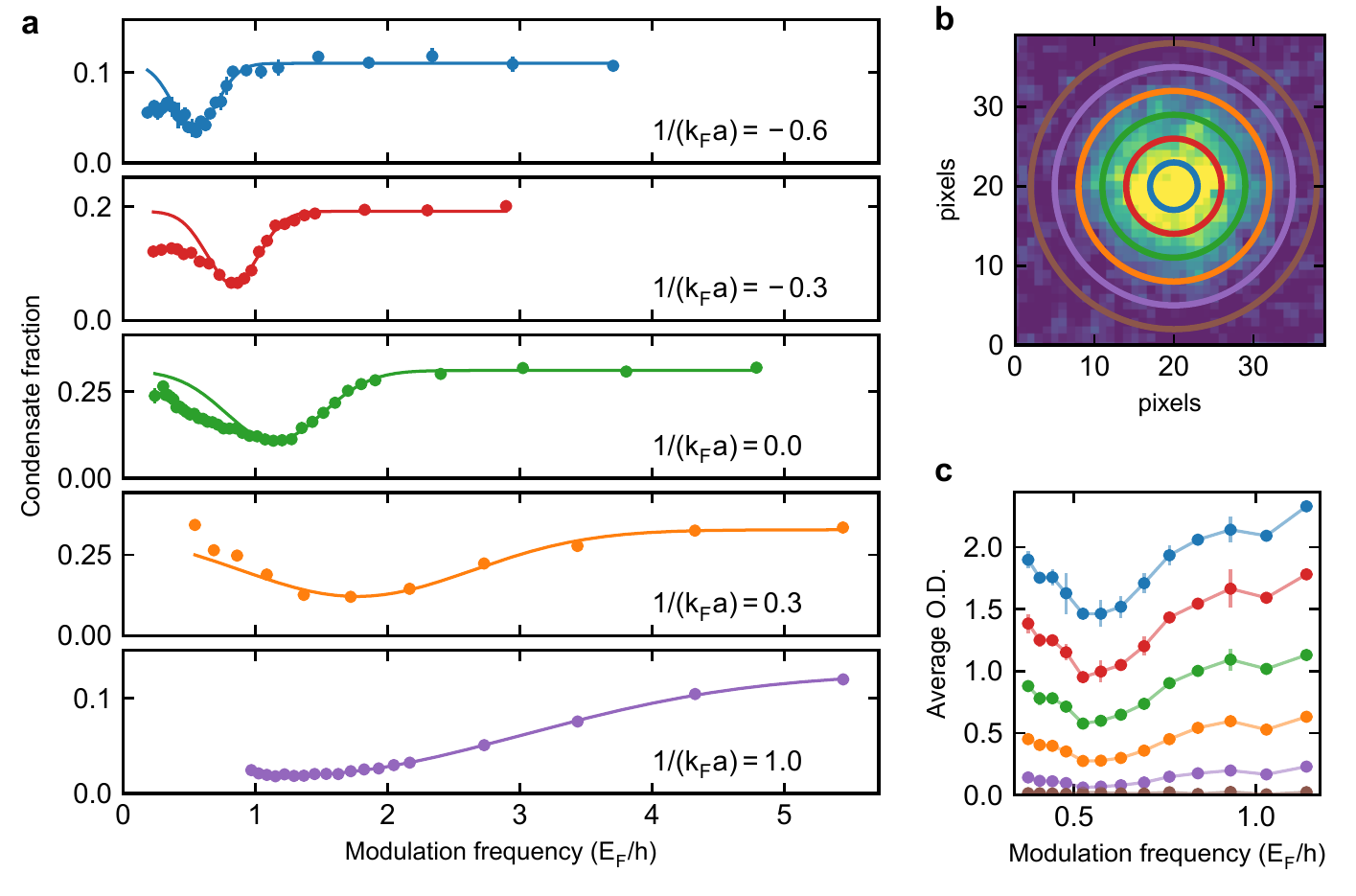}
 \caption{Excitation spectra of the Higgs mode. (a) Excitation spectra of the Higgs mode for different interaction strengths, $1/(k_F a)$, as labeled in the figure. The different levels of background condensate fraction are due to the different $1/(k_F a)$. The solid lines shows the Gaussian fit to the high frequency side of the spectra. The error bars show the standard deviation of approximately four measurements. (b) Time-of-flight image of the condensate with the thermal background subtracted at $1/(k_F a) \approx -0.43$. Rings indicate momentum intervals of $0.02\,k_F$. (c) Momentum-resolved analysis of the Higgs excitation inside the condensate by averaging the optical density in the color-coded rings in (b) for different modulation frequency. The resonance occurs at the same modulation frequency for all momenta. }
\end{figure}

In the experiment, we search for the Higgs mode by measuring the energy absorption spectrum of the fermionic superfluid in the $\ket{1,2}$ states for different interactions.  Using $\Omega_R$ and $\delta$ as adjustable parameters, we dress the $\ket{2}$ state by the $\ket{3}$ state with adjustable modulation frequency given by the effective Rabi frequency. We choose a drive frequency in the single--particle excitation gap. For our experiments we measure the modulation frequency $\Omega_\text{mod}$ and amplitude $\alpha$ of the time-dependent population of the $\ket{3}$ state (for calibration, see Methods and Supplementary Figure). We then use a constant excitation amplitude $p_{\ket{3}}\simeq0.5\%$ and apply the modulation for a fixed period of 30\,ms.  After the excitation we conduct a rapid magnetic field sweep onto the molecular side of the Feshbach resonance and convert Cooper pairs into dimers and measure the condensate fraction of the molecular condensate in time-of-flight imaging. The change in the condensate fraction provides us with a sensitive measure of the excitation of modes in the quantum gas.  
In Figure 3a we plot the measured spectra as a function of the modulation frequency for different values of $1/(k_Fa)$. On the BCS-side of the Feshbach resonance up to unitarity, $1/(k_Fa) < 0$, we observe clear resonances for which the condensate fraction reduces, signaling the excitation of a well defined mode.  For $1/(k_Fa) > 0$, the energy absorption peak is gradually washed out and broadens significantly. Far on the BEC side, for $1/(k_Fa) \simeq 1$, we cannot observe a resonance  and conclude that the Higgs mode is absent. The resonances generally exhibit an asymmetric line shape, which we fit with a Gaussian to the high-frequency side in order to extract the peak position and width. A contribution to the asymmetric peak shape stems from the momentum-dependence of the effective Rabi frequencies $\Omega_{R,k}^\prime$. As indicated in Figures 2a and b, the effective detuning (and hence the modulation frequency) varies with increasing momentum $k$. Therefore, a resonant excitation at the Higgs mode frequency can be achieved for high momenta $k$ even though  for low momenta the modulation frequency is below the resonant excitation.

In order to demonstrate the collective mode nature of our resonance, we perform a number of checks. Firstly, we verify that the excitation resonance frequency  (to within 4\%) and  shape is independent of the modulation strength in the range of $0.001<\alpha<0.2$ and modulation duration between $\tau=0.5\,$ms and $\tau=30$\,ms (for the definition of $\alpha$, see Methods and Supplementary Figure). Secondly, we have confirmed that the observed  resonance peak is not caused by single-particle excitations by measuring the excitation probability to the $\ket{3}$ state vs. modulation frequency $\Omega_\text{mod}$ and finding a  featureless broad  spectrum. This and the following checks have been performed with a modulation amplitude of $\alpha=0.2$ and a modulation time of $0.5\,$ms, which is much shorter than the trap period of $\sim 5$\,ms. Hence the measurement is insensitive to thermalization effects and/or density redistribution within the cloud. Thirdly, we check  the momentum-dependence of the resonance. After the modulation, we perform a time-of-flight expansion for a period of 15\,ms, which is approximately a quarter period of the residual harmonic potential during ballistic expansion. This procedure maps the initial momentum states to positions in the absorption image \cite{Ries2015}. We analyse the detected condensate density in momentum intervals of 0.02\,$k_F$ and find that the excitation resonance is at the same frequency for all momentum intervals, see Figure 3b and 3c. Finally, we have searched for possible quasiparticle excitations resulting from our interaction modulation by employing standard rf spectroscopy \cite{Stewart2008,Feld2011,Schirotzek2008} after the interaction modulation. The spectra only show a very weak and broad background independent of the modulation frequency. This behaviour is not unexpected since the contribution of quasi-particle excitations is smeared out in the presence of a trap as we  confirmed numerically using the local-density approximation. 

\begin{figure}
 \includegraphics[width=.5\columnwidth,clip=true]{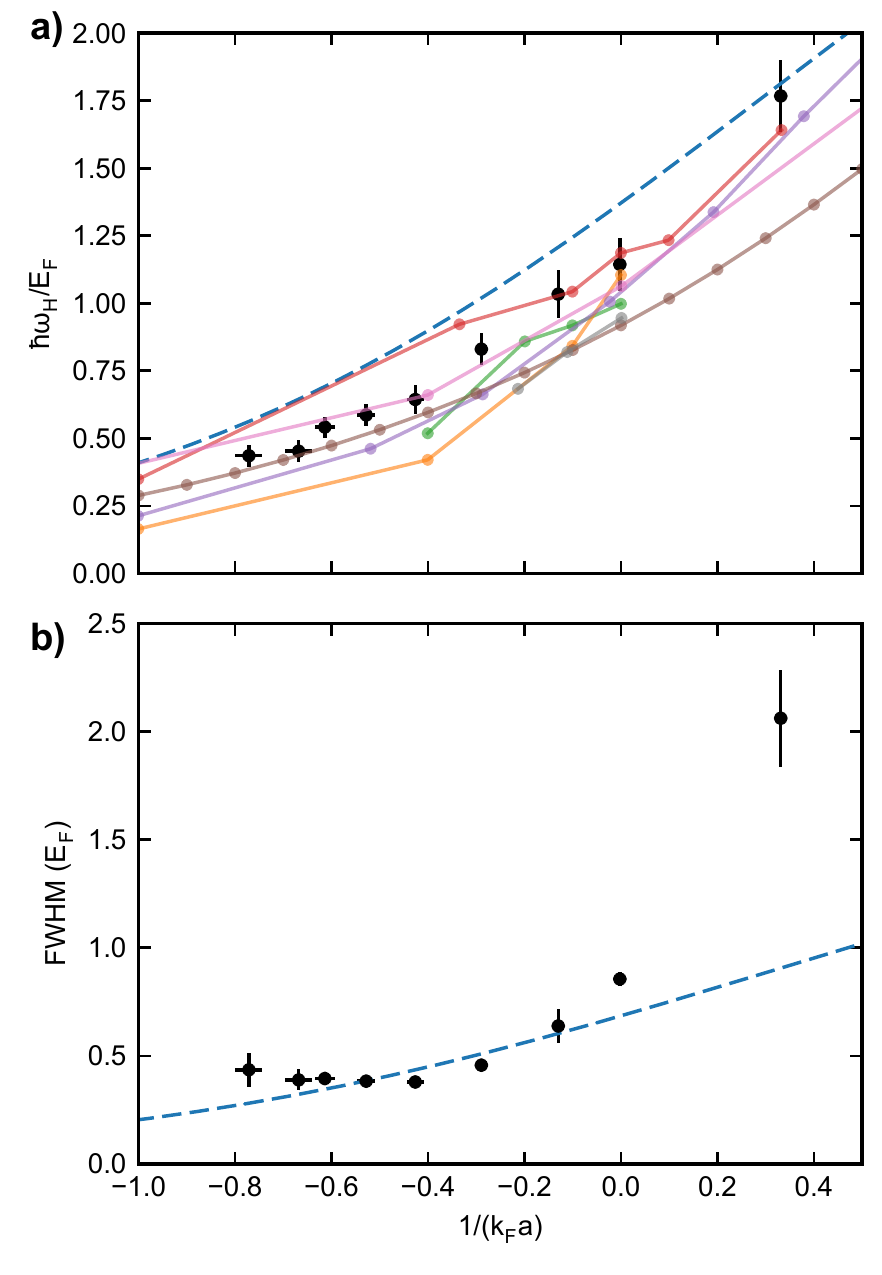}
 \caption{Observation of the Higgs mode. (a) Measured peak positions of the energy absorption spectra (black circles). For comparison we show numerical simulations of the gap parameter multiplied by 2: BCS mean field theory (blue dashed line), ref. \cite{Chang2004} (red), ref. \cite{Gezerlis2008} (green), ref. \cite{Bulgac2008} (orange), ref. \cite{Chen2016} (purple), ref. \cite{Haussmann2007} (brown), ref. \cite{Pieri2004} (pink). The grey symbols show the experimental data of ref. \cite{Hoinka2017}. (b) Measured fullwidth at half maximum (FWHM) of the absorption peaks (black circles). For comparison, the BCS mean field theory gap is also shown (blue dashed line). The error bars in a and b represent the standard errors.}
\end{figure}

In Figure 4a, we plot the position of the fitted peak of the energy absorption spectra vs. the interaction parameter $1/(k_Fa)$ evaluated at the center of the sample. It has been suggested\cite{Scott2012} that the Higgs mode frequency is close to twice the superconducting gap in the BCS/BEC crossover and can therefore be used as an approximative measure of the gap. In the crossover regime, the exact value of $|\Delta|$ is yet unknown and both experiments and numerical calculations are challenging. We compare our data to gap measurement using different methods\cite{Schirotzek2008,Hoinka2017} and  several numerical calculations\cite{Chang2004,Gezerlis2008,Bulgac2008,Chen2016, Haussmann2007,Pieri2004}. As compared to the previous experimental results, our extracted value is somewhat larger. We note that previous gap measurements rely on fitting the onset of a spectral feature whereas our method is based on fitting a Gaussian to a slightly skewed spectral feature and both methodologies could be susceptible to small systematic uncertainties. An upper bound is provided by the theoretical result of mean-field theory (dashed line), which is known to overestimate the superconducting gap.

In Figure 4b, we plot the full-width-at-half-maximum of the Gaussian fits to the energy absorption peaks. Utilizing the BCS model for the momentum-dependent excitation discussed above, we estimate the width of our excitation resonance to be of order $\Delta$, see Supplementary Figure.  Hence, we cannot directly interpret the linewidth of our spectra as the decay rate of the Higgs mode but only as lower limit of the lifetime. On the BCS side of the resonance we find good agreement with  our model and towards the BEC side the measured width far exceeds the prediction, indicating that the Higgs mode becomes strongly broadened, for example, due to the violation of particle-hole symmetry resulting in a  decay into Goldstone modes\cite{Ohashi2003,Pekker2015,Liu2016,Han2016}. Extending, in the future,  our novel experimental scheme with a better momentum resolution will provide a route to finally explore the decay mechanisms of the Higgs mode, the understanding of which is a cornerstone in both high-energy particle physics and condensed-matter physics.

\begin{addendum}
\item[Acknowledgements] We thank E. Demler, W. Zwerger and M. Zwierlein for fruitful discussion. This work has been supported by BCGS, the Alexander-von-Humboldt Stiftung, ERC (grant 616082 and 648166), DFG (SFB/TR 185 project B4),  ITN COMIQ and Studienstiftung des Deutschen Volkes.
\item[Competing Interests] The authors declare that they have no competing financial interests.
\item[Author Contributions]  The study was conceived by C.K. and M.K.; the experimental setup was designed and constructed by A.B., T.H., K.G. and M.K., data taking was performed by A.B., T.H., K.G., M.L.; data analysis was performed by T.H., numerical modelling and analysis was performed by J.K., J.-S.B. and C.K.; the manuscript was written by C.K. and M.K. with contributions from all coauthors. 
\item [Data Availability Statement] The data that support the plots within this paper and other findings of this study are available from the corresponding author upon reasonable request.”
\item[Correspondence] Correspondence and requests for materials should be addressed to M.K.~(email: michael.koehl@uni-bonn.de) or K.G. (email: kuiyi@physik.uni-bonn.de).
\end{addendum}

\begin{methods}

\section{Preparation}
Using standard techniques of laser cooling and sympathetic cooling in a mixture with Sodium atoms in a magnetic trap, we prepare $\sim 5 \times 10^7$ cold fermionic Lithium atoms in a crossed-beam optical dipole trap of wavelength 1070\,nm in an equal mixture of the two lowest hyperfine states $\ket{1}$ and $\ket{2}$. Using subsequent evaporative cooling in a homogeneous magnetic field of 795\,G, in immediate vicinity of the Feshbach resonance at 834\,G, we produce a  condensate in the BCS/BEC crossover regime with a temperature of $T/T_F=0.07\pm 0.02$. After preparation of the fermionic superfluid, the magnetic offset field is adiabatically adjusted in the range between 740\,G and 1000\,G in order to control the interaction parameter $1/(k_Fa)$ in the range of $-0.8<1/(k_Fa)<1$, i.e. across the whole BCS/BEC crossover region. 

\section{Calibration of spectroscopy and analysis}
We experimentally calibrate the modulation frequency and amplitude to take into account energy shifts owing to interaction effects of the initial and final states and the efficiency of the rf antenna setup. To this end, we drive Rabi oscillations with set values of detuning $\delta$ and power and measure the population $p_{\ket{3}}$ as a function of time during the rf drive $p_{\ket{3}}=\alpha \sin^2(\Omega_\text{mod}t/2)$. This provides us with a direct measurement of the modulation frequency and amplitude. In order to model the data, we assume a Lorentzian line shape $\alpha=\frac{\Omega_R^2}{\Omega_R^2+(\delta-\delta_0)^2}$, however, we allow for a frequency shift $\delta_0(k_Fa)$ by which the detuning $\delta$ is corrected as compared to the Zeeman-energy resonance of the free atom. The fit parameter $\delta_0$ absorbs the effects of interactions in the final state of the spectroscopy, the condensation energy of the initial state, and the averaging of different momentum states and densities in the trap. Experimentally, the calibration is performed at a value of $\alpha\simeq 4\%$ for which we obtain agreement with the Lorentzian model to a few percent. 

We check for unpaired atoms in the $\ket{2}$ state for red and blue detuned radio-frequency with respect to $\delta_0$ as a result of the modulation. This has been achieved by rapidly ramping the field to 450\,G with approximately $4$\,G/$\mu$s, which allows for the detection of free atoms rather than paired atoms. In the case of a red-detuned radio-frequency modulation no enhancement of the signal of unpaired atoms could be observed over the whole range of modulation frequencies. However, blue detuned radio-frequency modulation increases significantly  the number of unpaired atoms due to single particle excitations to the continuum, see Supplementary Figure 1. 

Additionally, we vary the driving strength $\alpha$ and observe that the resonance position of the peak with respect to the modulation frequency does not vary, see Supplementary Figure 2.

\section{Theoretical modelling}
The experimental system can be described taking three different fermionic levels into account. Initially the system is prepared in a balanced mixture of states $\ket{1}$ and $\ket{2}$.  Since we are mainly interested in the excitation mechanism and for this mainly the presence of a difference in the interaction strength between states   ($\ket{1}$ and $\ket{2}$) and  ($\ket{1}$ and $\ket{3}$) is needed, we take here only the interaction between these two states into account and decouple this term within the s-wave BCS channel. Using the rotating wave approximation for the coupling between the states $\ket{2}$ and $\ket{3}$, we obtain the Hamiltonian

\begin{equation}
H = H_{BCS} + \sum_{{\bm{k}}} (\varepsilon_{{\bm{k}}}  -  \hbar \delta) n_{{\bm{k}},3} +  \frac{\hbar \Omega_{R}}{2} \sum_{{\bm{k}}} \Big( c^{\dagger}_{{\bm{k}}, 3} c_{{\bm{k}}, 2} + c^{\dagger}_{{\bm{k}}, 2} c_{{\bm{k}}, 3} \Big) \nonumber
\end{equation}
with
\begin{equation}
 H_{BCS}=  \sum_{{\bm k}} \varepsilon_{{\bm k}}  (n_{{\bm k}, 1} + n_{{\bm k},2}) + \sum_{{\bm k}} \left\{ \Delta^{*} c_{-{\bm k}, 2} c_{{\bm k}, 1} + \Delta c^{\dagger}_{{\bm k}, 1} c^{\dagger}_{-{\bm k}, 2} \right\}. 
\end{equation}
Here $\Delta= \frac{g}{V} \sum_{\bm k} \braket{c_{-{\bm k}, 2} c_{{\bm k}, 1}} $, $\Omega_R$ is the  Rabi-frequency, $g$ the interaction strength, $V$ the volume, and the momentum independent detuning is  $\hbar \delta = \hbar \omega_{rf} - (\varepsilon^{0}_{3} - \varepsilon^{0}_{2})$, where $\varepsilon^0_n$ is the bare energy for the state $n=2,3$ and $\varepsilon_k=\hbar^2 k^2/(2m)$ is the single-particle dispersion. We determine $g$ from the scattering length using the expression provided in Ref.~\cite{Ketterle2007}.

In order to determine the time evolution of the order parameter, we derive a closed set of equations for the expectation values
\begin{eqnarray}
\hbar\frac{\partial }{\partial t} \langle c_{-{\bm k},2} c_{{\bm k},1} \rangle& = & i  \{-2\epsilon_{{\bm k}} \langle c_{-{\bm k},2} c_{{\bm k},1} \rangle - \frac{\hbar \Omega_R}{2} \langle c_{-{\bm k}, 3} c_{{\bm k}, 1} \rangle + \Delta (n_{{\bm k},1} +n_{-{\bm k},2} - 1)  \}  \nonumber \\
\hbar\frac{\partial }{\partial t} \langle c_{-{\bm k}, 3} c_{{\bm k}, 1} \rangle& = & i  \{- \frac{\hbar \Omega_R}{2} \langle c_{-{\bm k},2} c_{{\bm k},1} \rangle -(2\epsilon_{{\bm k}} - \hbar \delta) \langle c_{-{\bm k}, 3} c_{{\bm k}, 1} \rangle + \Delta \langle c^{\dagger}_{-{\bm k}, 2} c_{-{\bm k},3} \rangle \}  \nonumber \\
\hbar\frac{\partial }{\partial t} \langle c^{\dagger}_{-{\bm k}, 2} c_{-{\bm k},3} \rangle& = & i  \{\Delta^{*}\langle c_{-{\bm k}, 3} c_{{\bm k}, 1} \rangle + \hbar \delta \langle c^{\dagger}_{-{\bm k}, 2} c_{-{\bm k},3} \rangle - \frac{\hbar \Omega_R}{2} (n_{-{\bm k},2} - n_{-{\bm k}, 3} )  \}  \nonumber \\
\hbar\frac{\partial }{\partial t} n_{{\bm k},1}& = & -2~\text{Im}(\Delta^*~\langle c_{-{{\bm k}},2} c_{{{\bm k}},1} \rangle) \nonumber \\
\hbar\frac{\partial }{\partial t} n_{-{\bm k},2}& = & - 2~\text{Im}(\Delta^*~\langle c_{-{{\bm k}},2} c_{{{\bm k}},1} \rangle) + \hbar~\Omega_R~\text{Im}(\langle c^\dagger_{-{{\bm k}},2} c_{-{{\bm k}},3} \rangle)    \nonumber \\
\hbar\frac{\partial }{\partial t} n_{-{\bm k}, 3}& = & - \hbar~\Omega_R~\text{Im}(\langle c^\dagger_{-{{\bm k}},2} c_{-{{\bm k}},3} \rangle)  \nonumber  ,
\end{eqnarray}
 where the number densities are defined as $n_{{\bm k},m} = \langle c^{\dagger}_{{\bm k}, m} c_{{\bm k},m} \rangle$ with $m=1,2,3$. We solve these equations numerically discretizing both time $t$ and momentum $k$ and using the self-consistency condition $\Delta = \frac{g}{V} \sum_{{\bm k}} \langle c_{-{\bm k},2} c_{{\bm k},1} \rangle$ at each time step ensuring both the convergence for the time-step $d t$ and the momentum spacing. Typical values taken are $dk/k_F=5 \times 10^{-4}$, $dt=5\times 10^{-4} \hbar/E_F$ and the cutoff for the momentum sum is  $E_c=100\,E_F$.

The momentum resolved spectral weight of the gap shown in Figure 2c is computed as
\begin{equation}
A_{{\bm k}}(\omega) = V/g \Bigg|\mathscr{F} \left\{ \Big| \Delta_{{\bm k}}(t)\Big| - \frac{1}{T}\int_{0}^{T} dt \Big|\Delta_{{\bm k}}(t) \Big|\right\}\Bigg|
\end{equation}
with the momentum-dependent order parameter $\Delta_{\bm k}=g/V \braket{c_{-{\bm k},2}c_{{\bm k},1}}$. We use $T=400 \hbar/E_F$ for the calculation.

\section{Time evolution of the population of state $\ket{3}$}
We compare the theoretical evolution of the population of atoms in state $| 3 \rangle$ (see Figure 1c) during the application of the rf dressing to the experimental results. Supplementary Figure 3 shows the population of the atoms in state $| 3 \rangle$ normalized to the initial atom number in state $| 1 \rangle$. The simulation and experiment were performed with the same effective modulation frequency, $\Omega_{\text{mod}}$, and maximum atom transfer. Both curves show damped oscillations of the population of state $\ket{3}$ with time. The initial time behaviour up to approximately three oscillations agrees well between theory and experiment which means that the dominant damping mechanism is due to a dephasing of the different momentum components. Afterwards the experimental results show a stronger damping  which we attribute to other damping mechanisms such as for example the presence of collisions in the experiment which are not considered in the theoretical description. 

\section{Time evolution of the condensate fraction}
We show the evolution of the condensate fraction during the application of the rf dressing in Extended Data Fig. 4. After different durations of the application, the drive time, the rapid mapping to the condensate fraction has been performed and the condensate fraction has been taken. The drive amplitude was chosen to be 0.05\%. As a response, an oscillation of the condensate fraction close to the expected Rabi frequency can be observered over several oscillation periods with an amplitude of the order of 0.05 \%. 

\section{Comparison of the experimental and theoretical spectra}

In Supplementary Figure 5, we show a comparison of the experimentally measured spectra with the theoretical simulation. In order to gain insight into the structure expected from the excitation scheme, we theoretically extract the weight of the Higgs excitation for different effective modulation frequencies and plot these in the lower panel of Supplementary Figure 5 for $1/(k_{\text{F}}a)=-0.63$.  To evaluate the area under the Higgs, for each momentum, we integrate around the Higgs excitation peak (shown in Figure 2c) and then sum over all momenta along the Higgs excitation line. Let us note that this procedure leads to the artefact that at high modulation frequencies still a non-vanishing contribution to the weight is found, which, however, can be attributed to the excitation of quasi-particles in a homogeneous system and would vanish in a trapped system as considered experimentally. 
More importantly, we see that even though the Higgs mode has a very sharp frequency (as shown in Fig. 2c) and therefore a long life-time, the resulting spectra show a much broader peak. The width of the peak is due to the excitation procedure. In particular, a resonant excitation of the Higgs mode is already possible if the effective modulation frequency lies below the sharp frequency of the Higgs mode, since then already some of the Rabi frequencies of the higher momentum components (compare stars in Fig. 2c) can resonantly excite the Higgs mode. Thus, the broadening of the spectral feature is mainly due to the particular excitation scheme and not a measurement of the life-time of the Higgs mode. Let us conclude by pointing out that the full width at half maximum in both the theoretical and the experimental spectra is approximately $|\Delta|$. 

\section{Local density approximation for the quasi-particle excitations}
To study the effect of the harmonic trapping on the quasi-particle excitations we performed a calculation of the system’s dynamics within the local-density approximation (LDA). Within LDA we treat points of different density as effectively homogeneous systems with rescaled interaction $1/[k_{F}(\mathbf{r})a]$, Fermi energy $E_{F}(\mathbf{r})$  and chemical potential consistent with the system's density profile. We assume the latter to be the profile for non-interacting fermions as typically the density profiles only changes slightly for the considered interactions. The time evolution of the superconducting order parameter of the homogeneous system is performed locally for each point in the trap and rescaled to give $\frac{\Delta(\mathbf{r}, t)}{E_{F}(\mathbf{r} = 0)}$. We then take the density-weighted average of its Fourier transform. It is important to note that the Higgs - due to its collective nature - cannot be treated in this formalism, so that we remove the Higgs peak in each Fourier transform by hand before we take the trap average. Integrating the resulting spectrum gives the “background excitation weight” (cf. Supplementary Figure 6). In contrast to the peaked quasi-particle structure of a homogeneous system, we find the trap averaged “background excitation weight” to be significantly broadened resulting in a featureless, broad background.

\end{methods}


\begin{thebibliography}{10}
\expandafter\ifx\csname url\endcsname\relax
  \def\url#1{\texttt{#1}}\fi
\expandafter\ifx\csname urlprefix\endcsname\relax\def\urlprefix{URL }\fi
\providecommand{\bibinfo}[2]{#2}
\providecommand{\eprint}[2][]{\url{#2}}

\bibitem{Higgs1964}
\bibinfo{author}{{H}iggs, P.~W.}
\newblock \bibinfo{title}{Broken symmetries and the masses of gauge bosons}.
\newblock \emph{\bibinfo{journal}{Phys. Rev. Lett.}}
  \textbf{\bibinfo{volume}{13}}, \bibinfo{pages}{508--509}
  (\bibinfo{year}{1964}).
\newblock \urlprefix\url{https://link.aps.org/doi/10.1103/PhysRevLett.13.508}.

\bibitem{Littlewood1981}
\bibinfo{author}{Littlewood, P.~B.} \& \bibinfo{author}{Varma, C.~M.}
\newblock \bibinfo{title}{Gauge-invariant theory of the dynamical interaction
  of charge density waves and superconductivity}.
\newblock \emph{\bibinfo{journal}{Phys. Rev. Lett.}}
  \textbf{\bibinfo{volume}{47}}, \bibinfo{pages}{811--814}
  (\bibinfo{year}{1981}).
\newblock \urlprefix\url{http://link.aps.org/doi/10.1103/PhysRevLett.47.811}.

\bibitem{Sooryakumar1980}
\bibinfo{author}{Sooryakumar, R.} \& \bibinfo{author}{Klein, M.~V.}
\newblock \bibinfo{title}{Raman scattering by superconducting-gap excitations
  and their coupling to charge-density waves}.
\newblock \emph{\bibinfo{journal}{Phys. Rev. Lett.}}
  \textbf{\bibinfo{volume}{45}}, \bibinfo{pages}{660--662}
  (\bibinfo{year}{1980}).
\newblock \urlprefix\url{http://link.aps.org/doi/10.1103/PhysRevLett.45.660}.

\bibitem{Matsunaga2013}
\bibinfo{author}{Matsunaga, R.} \emph{et~al.}
\newblock \bibinfo{title}{{H}iggs amplitude mode in the {BCS} superconductors
  {N}b$_{1-x}${T}i$_x${N} induced by terahertz pulse excitation}.
\newblock \emph{\bibinfo{journal}{Phys. Rev. Lett.}}
  \textbf{\bibinfo{volume}{111}}, \bibinfo{pages}{057002}
  (\bibinfo{year}{2013}).
\newblock
  \urlprefix\url{http://link.aps.org/doi/10.1103/PhysRevLett.111.057002}.

\bibitem{Sherman2015}
\bibinfo{author}{Sherman, D.} \emph{et~al.}
\newblock \bibinfo{title}{The {{H}iggs} mode in disordered superconductors
  close to a quantum phase transition}.
\newblock \emph{\bibinfo{journal}{Nature Physics}}
  \textbf{\bibinfo{volume}{11}}, \bibinfo{pages}{188--192}
  (\bibinfo{year}{2015}).

\bibitem{Pekker2015}
\bibinfo{author}{Pekker, D.} \& \bibinfo{author}{Varma, C.}
\newblock \bibinfo{title}{Amplitude/{H}iggs modes in condensed matter physics}.
\newblock \emph{\bibinfo{journal}{Annual Review of Condensed Matter Physics}}
  \textbf{\bibinfo{volume}{6}}, \bibinfo{pages}{269--297}
  (\bibinfo{year}{2015}).

\bibitem{Podolsky2011}
\bibinfo{author}{Podolsky, D.}, \bibinfo{author}{Auerbach, A.} \&
  \bibinfo{author}{Arovas, D.~P.}
\newblock \bibinfo{title}{Visibility of the amplitude ({H}iggs) mode in
  condensed matter}.
\newblock \emph{\bibinfo{journal}{Phys. Rev. B}} \textbf{\bibinfo{volume}{84}},
  \bibinfo{pages}{174522} (\bibinfo{year}{2011}).
\newblock \urlprefix\url{http://link.aps.org/doi/10.1103/PhysRevB.84.174522}.

\bibitem{Scott2012}
\bibinfo{author}{Scott, R.~G.}, \bibinfo{author}{Dalfovo, F.},
  \bibinfo{author}{Pitaevskii, L.~P.} \& \bibinfo{author}{Stringari, S.}
\newblock \bibinfo{title}{Rapid ramps across the {BEC-BCS} crossover: A route
  to measuring the superfluid gap}.
\newblock \emph{\bibinfo{journal}{Phys. Rev. A}} \textbf{\bibinfo{volume}{86}},
  \bibinfo{pages}{053604} (\bibinfo{year}{2012}).
\newblock \urlprefix\url{http://link.aps.org/doi/10.1103/PhysRevA.86.053604}.

\bibitem{Barlas2013}
\bibinfo{author}{Barlas, Y.} \& \bibinfo{author}{Varma, C.~M.}
\newblock \bibinfo{title}{Amplitude or {H}iggs modes in $d$-wave
  superconductors}.
\newblock \emph{\bibinfo{journal}{Phys. Rev. B}} \textbf{\bibinfo{volume}{87}},
  \bibinfo{pages}{054503} (\bibinfo{year}{2013}).
\newblock \urlprefix\url{https://link.aps.org/doi/10.1103/PhysRevB.87.054503}.

\bibitem{Liu2016}
\bibinfo{author}{Liu, B.}, \bibinfo{author}{Zhai, H.} \&
  \bibinfo{author}{Zhang, S.}
\newblock \bibinfo{title}{Evolution of the {H}iggs mode in a fermion superfluid
  with tunable interactions}.
\newblock \emph{\bibinfo{journal}{Phys. Rev. A}} \textbf{\bibinfo{volume}{93}},
  \bibinfo{pages}{033641} (\bibinfo{year}{2016}).
\newblock \urlprefix\url{http://link.aps.org/doi/10.1103/PhysRevA.93.033641}.

\bibitem{Han2016}
\bibinfo{author}{Han, X.}, \bibinfo{author}{Liu, B.} \& \bibinfo{author}{Hu,
  J.}
\newblock \bibinfo{title}{Observability of {H}iggs mode in a system without
  {L}orentz invariance}.
\newblock \emph{\bibinfo{journal}{Phys. Rev. A}} \textbf{\bibinfo{volume}{94}},
  \bibinfo{pages}{033608} (\bibinfo{year}{2016}).
\newblock \urlprefix\url{http://link.aps.org/doi/10.1103/PhysRevA.94.033608}.

\bibitem{Littlewood1982}
\bibinfo{author}{Littlewood, P.~B.} \& \bibinfo{author}{Varma, C.~M.}
\newblock \bibinfo{title}{Amplitude collective modes in superconductors and
  their coupling to charge-density waves}.
\newblock \emph{\bibinfo{journal}{Phys. Rev. B}} \textbf{\bibinfo{volume}{26}},
  \bibinfo{pages}{4883--4893} (\bibinfo{year}{1982}).
\newblock \urlprefix\url{http://link.aps.org/doi/10.1103/PhysRevB.26.4883}.

\bibitem{Ruegg2008}
\bibinfo{author}{R\"uegg, C.} \emph{et~al.}
\newblock \bibinfo{title}{Quantum magnets under pressure: Controlling
  elementary excitations in {T}l{C}u{C}l$_{3}$}.
\newblock \emph{\bibinfo{journal}{Phys. Rev. Lett.}}
  \textbf{\bibinfo{volume}{100}}, \bibinfo{pages}{205701}
  (\bibinfo{year}{2008}).
\newblock
  \urlprefix\url{https://link.aps.org/doi/10.1103/PhysRevLett.100.205701}.

\bibitem{Halperin1990}
\bibinfo{author}{Halperin, W.} \& \bibinfo{author}{Varoquax, E.}
\newblock \bibinfo{title}{Order-parameter collective modes in superfluid
  {3He}}.
\newblock In \bibinfo{editor}{Halperin, W.} \& \bibinfo{editor}{Pitaevskii, L.}
  (eds.) \emph{\bibinfo{booktitle}{Helium Three}}, \bibinfo{pages}{353--522}
  (\bibinfo{publisher}{Elsevier Science Publishers}, \bibinfo{year}{1990}).

\bibitem{Bissbort2011}
\bibinfo{author}{Bissbort, U.} \emph{et~al.}
\newblock \bibinfo{title}{Detecting the amplitude mode of strongly interacting
  lattice bosons by {B}ragg scattering}.
\newblock \emph{\bibinfo{journal}{Phys. Rev. Lett.}}
  \textbf{\bibinfo{volume}{106}}, \bibinfo{pages}{205303}
  (\bibinfo{year}{2011}).
\newblock
  \urlprefix\url{http://link.aps.org/doi/10.1103/PhysRevLett.106.205303}.

\bibitem{Endres2012}
\bibinfo{author}{Endres, M.} \emph{et~al.}
\newblock \bibinfo{title}{The {‘{H}iggs’} amplitude mode at the
  two-dimensional superfluid/{M}ott insulator transition}.
\newblock \emph{\bibinfo{journal}{Nature}} \textbf{\bibinfo{volume}{487}},
  \bibinfo{pages}{454--458} (\bibinfo{year}{2012}).

\bibitem{Hoang2016}
\bibinfo{author}{Hoang, T.~M.} \emph{et~al.}
\newblock \bibinfo{title}{Adiabatic quenches and characterization of amplitude
  excitations in a continuous quantum phase transition}.
\newblock \emph{\bibinfo{journal}{Proceedings of the National Academy of
  Sciences}} \textbf{\bibinfo{volume}{113}}, \bibinfo{pages}{9475--9479}
  (\bibinfo{year}{2016}).
\newblock \urlprefix\url{http://www.pnas.org/content/113/34/9475.abstract}.
\newblock \eprint{http://www.pnas.org/content/113/34/9475.full.pdf}.

\bibitem{Leonard2017}
\bibinfo{author}{Leonard, J.}, \bibinfo{author}{Morales, A.},
  \bibinfo{author}{Zupancic, P.}, \bibinfo{author}{Donner, T.} \&
  \bibinfo{author}{Esslinger, T.}
\newblock \bibinfo{title}{Monitoring and manipulating {H}iggs and {G}oldstone
  modes in a supersolid quantum gas}.
\newblock \emph{\bibinfo{journal}{Science}} \textbf{\bibinfo{volume}{358}},
  \bibinfo{pages}{1415--1418} (\bibinfo{year}{2017}).

\bibitem{Yuzbashyan2006}
\bibinfo{author}{Yuzbashyan, E.~A.} \& \bibinfo{author}{Dzero, M.}
\newblock \bibinfo{title}{Dynamical vanishing of the order parameter in a
  fermionic condensate}.
\newblock \emph{\bibinfo{journal}{Phys. Rev. Lett.}}
  \textbf{\bibinfo{volume}{96}}, \bibinfo{pages}{230404}
  (\bibinfo{year}{2006}).
\newblock
  \urlprefix\url{http://link.aps.org/doi/10.1103/PhysRevLett.96.230404}.

\bibitem{Hannibal2015}
\bibinfo{author}{Hannibal, S.} \emph{et~al.}
\newblock \bibinfo{title}{Quench dynamics of an ultracold {F}ermi gas in the
  {BCS} regime: Spectral properties and confinement-induced breakdown of the
  {H}iggs mode}.
\newblock \emph{\bibinfo{journal}{Phys. Rev. A}} \textbf{\bibinfo{volume}{91}},
  \bibinfo{pages}{043630} (\bibinfo{year}{2015}).
\newblock \urlprefix\url{http://link.aps.org/doi/10.1103/PhysRevA.91.043630}.

\bibitem{Greiner2005}
\bibinfo{author}{Greiner, M.}, \bibinfo{author}{Regal, C.~A.} \&
  \bibinfo{author}{Jin, D.~S.}
\newblock \bibinfo{title}{Probing the excitation spectrum of a {F}ermi gas in
  the {BCS-BEC} crossover regime}.
\newblock \emph{\bibinfo{journal}{Phys. Rev. Lett.}}
  \textbf{\bibinfo{volume}{94}}, \bibinfo{pages}{070403}
  (\bibinfo{year}{2005}).
\newblock
  \urlprefix\url{https://link.aps.org/doi/10.1103/PhysRevLett.94.070403}.

\bibitem{Chin2004}
\bibinfo{author}{Chin, C.} \emph{et~al.}
\newblock \bibinfo{title}{Observation of the pairing gap in a strongly
  interacting {{F}ermi} gas}.
\newblock \emph{\bibinfo{journal}{Science}} \textbf{\bibinfo{volume}{305}},
  \bibinfo{pages}{1128--1130} (\bibinfo{year}{2004}).

\bibitem{Ketterle2007}
\bibinfo{author}{Ketterle, W.} \& \bibinfo{author}{Zwierlein, M.~W.}
\newblock \bibinfo{title}{Making, probing and understanding ultracold {F}ermi
  gases}.
\newblock \emph{\bibinfo{journal}{Proceedings of the International School of
  Physics "Enrico Fermi"}} \textbf{\bibinfo{volume}{164}},
  \bibinfo{pages}{95--287} (\bibinfo{year}{2007}).

\bibitem{Stewart2008}
\bibinfo{author}{Stewart, J.~T.}, \bibinfo{author}{Gaebler, J.~P.} \&
  \bibinfo{author}{Jin, D.~S.}
\newblock \bibinfo{title}{Using photoemission spectroscopy to probe a strongly
  interacting {F}ermi gas}.
\newblock \emph{\bibinfo{journal}{Nature}} \textbf{\bibinfo{volume}{454}},
  \bibinfo{pages}{744--747} (\bibinfo{year}{2008}).

\bibitem{Feld2011}
\bibinfo{author}{Feld, M.}, \bibinfo{author}{Fr{\"o}hlich, B.},
  \bibinfo{author}{Vogt, E.}, \bibinfo{author}{Koschorreck, M.} \&
  \bibinfo{author}{K{\"o}hl, M.}
\newblock \bibinfo{title}{Observation of a pairing pseudogap in a
  two-dimensional {F}ermi gas}.
\newblock \emph{\bibinfo{journal}{Nature}} \textbf{\bibinfo{volume}{480}},
  \bibinfo{pages}{75--78} (\bibinfo{year}{2011}).

\bibitem{Bruun2002}
\bibinfo{author}{Bruun, G.~M.}
\newblock \bibinfo{title}{Low-energy monopole modes of a trapped atomic {F}ermi
  gas}.
\newblock \emph{\bibinfo{journal}{Phys. Rev. Lett.}}
  \textbf{\bibinfo{volume}{89}}, \bibinfo{pages}{263002}
  (\bibinfo{year}{2002}).
\newblock
  \urlprefix\url{https://link.aps.org/doi/10.1103/PhysRevLett.89.263002}.

\bibitem{Korolyuk2011}
\bibinfo{author}{Korolyuk, A.}, \bibinfo{author}{Kinnunen, J.~J.} \&
  \bibinfo{author}{T\"orm\"a, P.}
\newblock \bibinfo{title}{Density response of a trapped fermi gas: A crossover
  from the pair vibration mode to the goldstone mode}.
\newblock \emph{\bibinfo{journal}{Phys. Rev. A}} \textbf{\bibinfo{volume}{84}},
  \bibinfo{pages}{033623} (\bibinfo{year}{2011}).
\newblock \urlprefix\url{https://link.aps.org/doi/10.1103/PhysRevA.84.033623}.

\bibitem{Korolyuk2014}
\bibinfo{author}{Korolyuk, A.}, \bibinfo{author}{Kinnunen, J.~J.} \&
  \bibinfo{author}{T\"orm\"a, P.}
\newblock \bibinfo{title}{Collective excitations of a trapped fermi gas at
  finite temperature}.
\newblock \emph{\bibinfo{journal}{Phys. Rev. A}} \textbf{\bibinfo{volume}{89}},
  \bibinfo{pages}{013602} (\bibinfo{year}{2014}).
\newblock \urlprefix\url{https://link.aps.org/doi/10.1103/PhysRevA.89.013602}.

\bibitem{Tokimoto2017}
\bibinfo{author}{Tokimoto, J.}, \bibinfo{author}{Tsuchiya, S.} \&
  \bibinfo{author}{Nikuni, T.}
\newblock \bibinfo{title}{{H}iggs mode in a trapped superfluid {F}ermi gas}.
\newblock \emph{\bibinfo{journal}{Journal of Low Temperature Physics}}
  \bibinfo{pages}{1--6} (\bibinfo{year}{2017}).
\newblock \urlprefix\url{http://dx.doi.org/10.1007/s10909-017-1766-2}.

\bibitem{Ries2015}
\bibinfo{author}{Ries, M.~G.} \emph{et~al.}
\newblock \bibinfo{title}{Observation of pair condensation in the quasi-2{D}
  {BEC-BCS} crossover}.
\newblock \emph{\bibinfo{journal}{Phys. Rev. Lett.}}
  \textbf{\bibinfo{volume}{114}}, \bibinfo{pages}{230401}
  (\bibinfo{year}{2015}).
\newblock
  \urlprefix\url{https://link.aps.org/doi/10.1103/PhysRevLett.114.230401}.

\bibitem{Schirotzek2008}
\bibinfo{author}{Schirotzek, A.}, \bibinfo{author}{Shin, Y.-i.},
  \bibinfo{author}{Schunck, C.~H.} \& \bibinfo{author}{Ketterle, W.}
\newblock \bibinfo{title}{Determination of the superfluid gap in atomic {F}ermi
  gases by quasiparticle spectroscopy}.
\newblock \emph{\bibinfo{journal}{Phys. Rev. Lett.}}
  \textbf{\bibinfo{volume}{101}}, \bibinfo{pages}{140403}
  (\bibinfo{year}{2008}).
\newblock
  \urlprefix\url{https://link.aps.org/doi/10.1103/PhysRevLett.101.140403}.

\bibitem{Hoinka2017}
\bibinfo{author}{Hoinka, S.} \emph{et~al.}
\newblock \bibinfo{title}{Goldstone mode and pair-breaking excitations in
  atomic {F}ermi superfluids}.
\newblock \emph{\bibinfo{journal}{Nature Physics}}
  \textbf{\bibinfo{volume}{13}}, \bibinfo{pages}{943–--946}
  (\bibinfo{year}{2017}).

\bibitem{Chang2004}
\bibinfo{author}{Chang, S.~Y.}, \bibinfo{author}{Pandharipande, V.~R.},
  \bibinfo{author}{Carlson, J.} \& \bibinfo{author}{Schmidt, K.~E.}
\newblock \bibinfo{title}{Quantum {M}onte {C}arlo studies of superfluid {F}ermi
  gases}.
\newblock \emph{\bibinfo{journal}{Phys. Rev. A}} \textbf{\bibinfo{volume}{70}},
  \bibinfo{pages}{043602} (\bibinfo{year}{2004}).
\newblock \urlprefix\url{https://link.aps.org/doi/10.1103/PhysRevA.70.043602}.

\bibitem{Gezerlis2008}
\bibinfo{author}{Gezerlis, A.} \& \bibinfo{author}{Carlson, J.}
\newblock \bibinfo{title}{Strongly paired fermions: Cold atoms and neutron
  matter}.
\newblock \emph{\bibinfo{journal}{Phys. Rev. C}} \textbf{\bibinfo{volume}{77}},
  \bibinfo{pages}{032801} (\bibinfo{year}{2008}).
\newblock \urlprefix\url{https://link.aps.org/doi/10.1103/PhysRevC.77.032801}.

\bibitem{Bulgac2008}
\bibinfo{author}{Bulgac, A.}, \bibinfo{author}{Drut, J.~E.} \&
  \bibinfo{author}{Magierski, P.}
\newblock \bibinfo{title}{Quantum {M}onte {C}arlo simulations of the {BCS-BEC}
  crossover at finite temperature}.
\newblock \emph{\bibinfo{journal}{Phys. Rev. A}} \textbf{\bibinfo{volume}{78}},
  \bibinfo{pages}{023625} (\bibinfo{year}{2008}).
\newblock \urlprefix\url{https://link.aps.org/doi/10.1103/PhysRevA.78.023625}.

\bibitem{Chen2016}
\bibinfo{author}{Chen, Q.}
\newblock \bibinfo{title}{Effect of the particle-hole channel on
  {BCS}-–{B}ose--{E}instein condensation crossover in atomic {F}ermi gases}.
\newblock \emph{\bibinfo{journal}{Scientific Reports}}
  \textbf{\bibinfo{volume}{6}}, \bibinfo{pages}{25772} (\bibinfo{year}{2016}).

\bibitem{Haussmann2007}
\bibinfo{author}{Haussmann, R.}, \bibinfo{author}{Rantner, W.},
  \bibinfo{author}{Cerrito, S.} \& \bibinfo{author}{Zwerger, W.}
\newblock \bibinfo{title}{Thermodynamics of the {BCS-BEC} crossover}.
\newblock \emph{\bibinfo{journal}{Phys. Rev. A}} \textbf{\bibinfo{volume}{75}},
  \bibinfo{pages}{023610} (\bibinfo{year}{2007}).
\newblock \urlprefix\url{https://link.aps.org/doi/10.1103/PhysRevA.75.023610}.

\bibitem{Pieri2004}
\bibinfo{author}{Pieri, P.}, \bibinfo{author}{Pisani, L.} \&
  \bibinfo{author}{Strinati, G.~C.}
\newblock \bibinfo{title}{{BCS-BEC} crossover at finite temperature in the
  broken-symmetry phase}.
\newblock \emph{\bibinfo{journal}{Phys. Rev. B}} \textbf{\bibinfo{volume}{70}},
  \bibinfo{pages}{094508} (\bibinfo{year}{2004}).
\newblock \urlprefix\url{https://link.aps.org/doi/10.1103/PhysRevB.70.094508}.

\bibitem{Ohashi2003}
\bibinfo{author}{Ohashi, Y.} \& \bibinfo{author}{Griffin, A.}
\newblock \bibinfo{title}{Superfluidity and collective modes in a uniform gas
  of {F}ermi atoms with a {F}eshbach resonance}.
\newblock \emph{\bibinfo{journal}{Phys. Rev. A}} \textbf{\bibinfo{volume}{67}},
  \bibinfo{pages}{063612} (\bibinfo{year}{2003}).
\newblock \urlprefix\url{https://link.aps.org/doi/10.1103/PhysRevA.67.063612}.

\end{thebibliography}
\end{document}